\documentclass[reprint,
 amsmath,amssymb,
 aps,
]{revtex4-2}

\usepackage{graphicx}
\usepackage{dcolumn}
\usepackage{bm}

\usepackage{xcolor}

\begin{document}

\preprint{APS/123-QED}

\title{Production of Intense Spin-Polarized Beams of Hydrogen Isotopes by Charge Transfer with High Density Raman-Pumped Alkali-Metal Vapors}

\author{Thad G. Walker}
 \email{tgwalker@wisc.edu}
 \author{Cary B. Forest}
\author{Deniz D. Yavuz}%
\affiliation{%
Dept. of Physics, University of Wisconsin-Madison
}%

\date{\today}

\begin{abstract}

It should be possible to generate multi-ampere spin-polarized beams of hydrogen isotopes  by repeated charge-transfer collisions in highly spin-polarized Cs vapor.  Estimates suggest that off-resonant Raman pumping with kW scale narrowband tunable light at 895 nm should be able to produce a 1 m long, 10 cm diameter volume of 80\% polarized Cs vapor.  The charge transfer collisions between the Cs and hydrogen result in a high nuclear spin-polarized negative ion beam that can be subsequently accelerated to high energy, neutralized, and be used to heat fusion plasmas with resulting increases in the fusion conversion efficiency.

\end{abstract}

\maketitle


\textit{Motivation and concept.--}
We argue that it should be possible to produce multi-ampere nuclear-spin-polarized negative ion beams of hydrogen isotopes by repeated $H^0-H^-$ charge-transfer collisions with high density spin-polarized vapors of the heavier alkali metals (K, Rb, Cs).  The required spin-polarized alkali-metal densities of $n\sim 10^{14}$ cm$^{-3}$ can be attained by off-resonant spontaneous Raman scattering from high-intensity laser light, largely avoiding depolarization by radiation trapping.  The transmitted Raman light can be recycled multiple times to produce the meter-scale spin-polarized column of polarized alkali-metal vapor required for production of nuclear-spin-polarized beams.  

Such beams are of substantial interest for boosting the efficiency of magnetically-confined fusion plasmas.  Broadly speaking, for fully nuclear-spin-polarized plasmas, the D-T fusion cross section proceeds largely through the $J=3/2$ collision channel; the $J=1/2$ channel is suppressed for spin-polarized nuclei, leading to a ~50\% enhancement in fusion rates \cite{Kulsrud1982FusionNuclei}.  The preservation of nuclear spin in a fusion plasma is a topic of great current interest \cite{Heidbrink2024AFuel}.  The relaxation mechanism of most obvious concern, transverse magnetic field fluctuations at the nuclear Larmor frequency $\omega_L$, is greatly suppressed by the collective motion of the electrons in magnetized plasma, so that fluctuations are primarily at the electron gyromagnetic frequency scale \cite{Kulsrud1986PhysicsPlasmas} $\omega_e\gg \omega_L$. A source that can deliver ampere-scale polarized neutral beams would therefore do more than improve a hypothetical reactor: it would provide the diagnostic tool needed to test spin survival in hot dense plasmas.


One approach to produce high energy, high current polarized neutral atom beams is to accelerate a polarized negative ion beam to the optimum $>100$ keV energies and then strip one electron.  Optically pumped polarized H/D$^-$ beams have been extensively developed for accelerator applications \cite{Zelenski2023PolarizedSources}, and achieve several mW of polarized H/D$^-$ current using charge exchange with optically pumped Rb vapor, followed by electron-nucleus spin transfer via the Sona effect\cite{Sona1967AD-}.  Such beams are however constrained by the need for low divergence in order to attain high efficiency coupling to polarized targets, and by relatively modest available laser power to optically pump the Rb vapor.  Plasma applications are much more forgiving regarding emittance, and for fusion applications it may be feasible to use much higher power lasers.  

In the following, we discuss the possibilities of producing high current polarized D$^-$ beams by using repeated charge transfer collisions with dense, spin-polarized Cs vapor for fusion applications. This builds on the original collisional pumping concepts of Anderson and co-workers \cite{Anderson1984}; the use of kW-scale lasers for off-resonant Raman pumping to polarize dense Cs vapor allows sufficiently large numbers of collisions to produce high current, high nuclear-spin-polarization negative ion beams.

\begin{figure}
    \centering
    \includegraphics[width=1\linewidth]{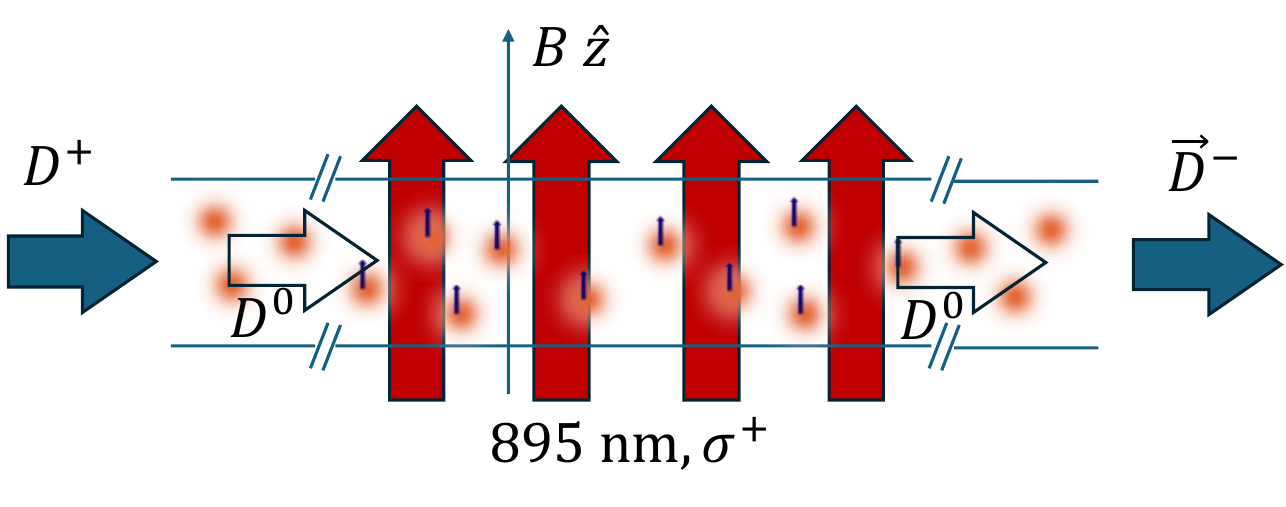}
    \caption{1 keV D$^+$ ions enter a long tube of transversely spin-polarized Cs vapor, produced by repeated zones of Raman pumping laser light.  Charge transfer pumping and hyperfine interactions produce a highly nuclear and electron spin-polarized D$^0$ current that is finally converted to $D^-$ for subsequent acceleration to high energy.}
    \label{fig:Schematic}
\end{figure}

For concreteness, we will specifically discuss using Cs vapor to produce a spin-polarized D$^-$ beam, but the principles apply to other combinations.  The concept is illustrated in Fig.~\ref{fig:Schematic}.  A multi-Ampere 1 keV D$^+$ beam enters a 1 m long, 10 cm diameter tube filled with transversely spin-polarized Cs vapor at 10$^{14}$ cm$^{-3}$ (~410 K), and a 10 G transverse magnetic field.  Within a few cm, the beam is neutralized.  

Electron spin-transfer collisions (u/d for Deuterium electron spin, arrows for Cs electron spin) $D^0_d+Cs\uparrow\rightarrow D^-+Cs^+$ produce D$^-$ from only the $D^0_d$ current due to the spin-singlet character of D$^-$.  The corresponding $D^0_u$ process is spin-forbidden. Subsequent electron detachment collisions $D^-+Cs\rightarrow D^0+Cs+e$ equally repopulate both $D^0_d$ and $D^0_u$ currents. Repetition of this charge-transfer pumping cycle results in the D$^0$ electron-spin polarization coming to equilibrium with the Cs electron-spin polarization. 

\begin{figure*}[t]
\centering
\includegraphics[width=0.4\textwidth]{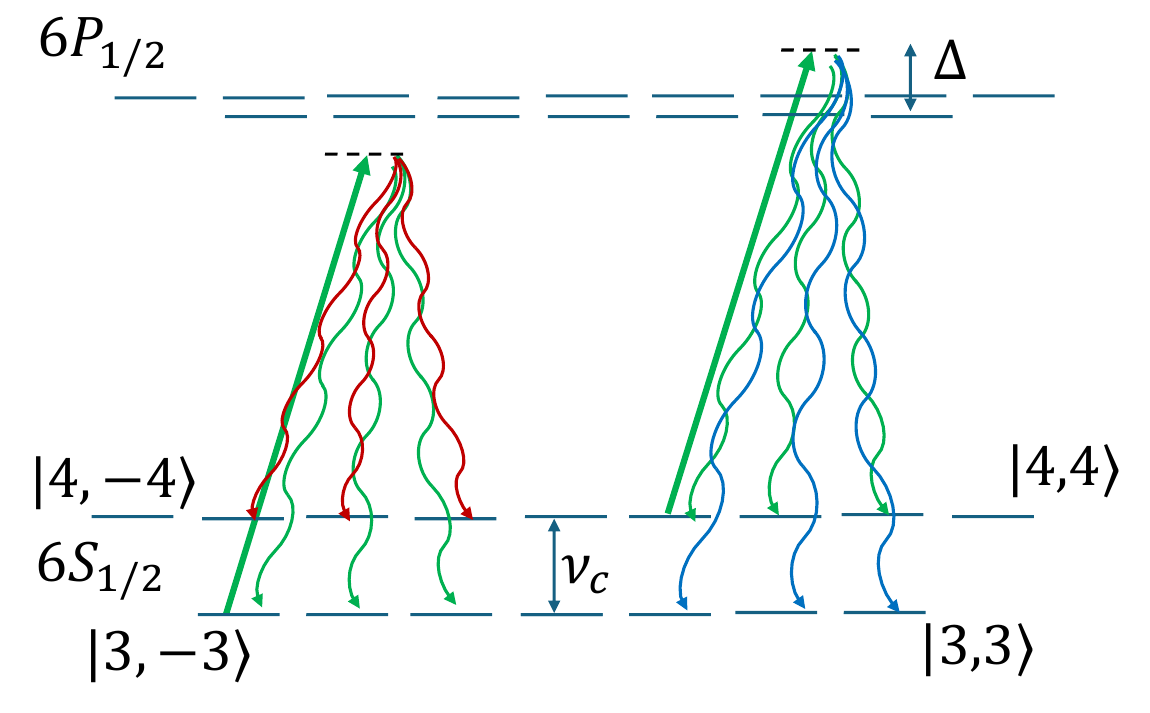}\hspace{0.025\textwidth}%
\includegraphics[width=0.4\textwidth]{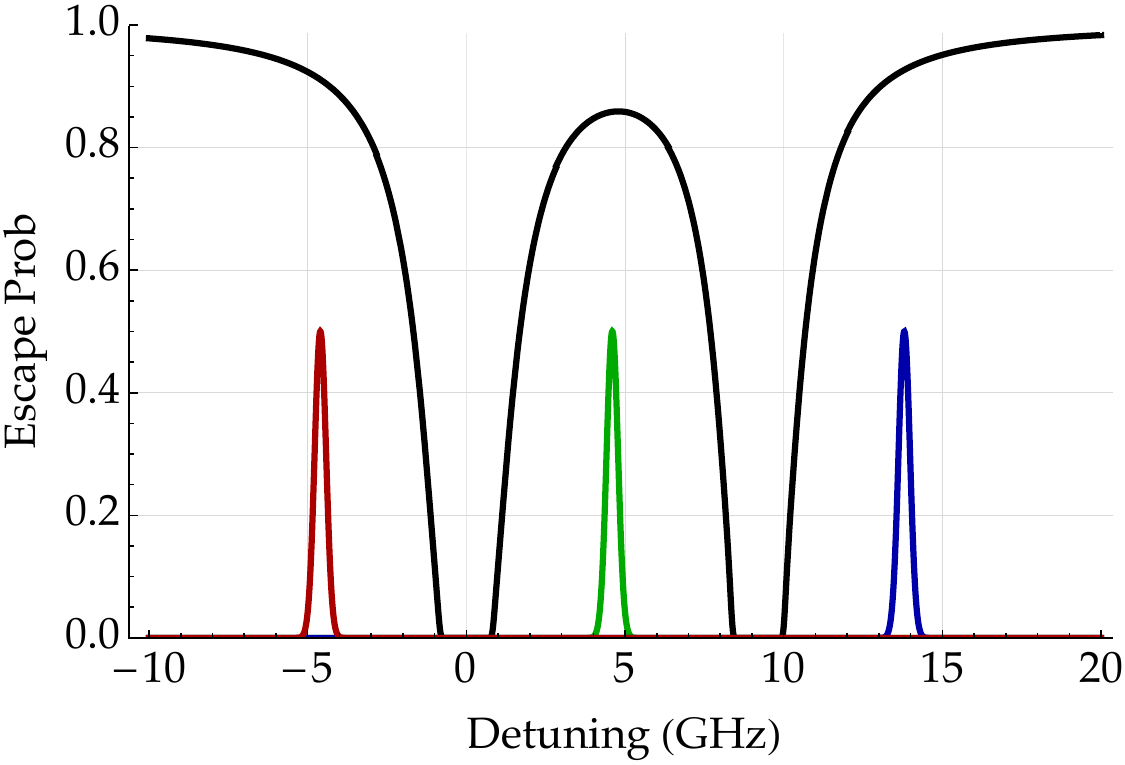}
\caption{Raman pumping.  (a) Circularly polarized light detuned by about half the Cs ground-state hyperfine splitting drives off-resonant Raman pumping toward stretched states. (b) Escape probability for emission of unpolarized light.  For $10^{14}\,\mathrm{cm}^{-3}$ Cs in a 10 cm diameter cell, the narrow Raman sidebands lie in high-escape-probability spectral windows.}
\label{fig:CsRaman}
\end{figure*}

As the charge transfer pumping repeats, nuclear spin-polarization is generated by the hyperfine interaction between the D$^0$ electron and nuclear spins. After many cycles, both electron and nuclear spin-polarizations come to equilibrium with the Cs electron-spin-polarization.

At the exit of the tube, the spin-polarized neutral beam is converted to a spin-polarized $D^-$ beam by spin-transfer with unpolarized Cs.  From there, it can be accelerated, then stripped and used for polarized neutral beam  heating of a plasma.

\textit{Raman pumping of high density Cs vapor--}
The fundamental challenge facing the production of collisionally thick spin-polarized alkali-metal targets is that such targets are extremely opaque to the resonance light normally used to  spin-polarize the atoms by optical pumping. Atom-atom collision cross sections at thermal energies and above are generally less than $10^{-14}$ cm$^2$, while Doppler-broadened light scattering cross sections are of order $10^{-11}$  cm$^2$.  Thus resonant optical depths exceed collisional depths by typically 3-4 orders of magnitude.

By far the most common means used to allow high density optical pumping is to introduce a molecular  gas, most often N$_2$, with rovibrational resonances close in energy to the excited state.  The need for the quenching time to be short compared to spontaneous emission lifetimes requires molecular densities in excess of ~10$^{16}$ cm$^{-3}$\cite{Rosenberry2007RadiationPressures,Lancor10c}.  Such high densities are clearly incompatible with atomic/ion beam propagation.

Consider instead optical pumping by off-resonant Raman scattering.  The specific case of Cs is illustrated in Fig.~\ref{fig:CsRaman}. The $\sigma^+$ pumping light is detuned from the $6S_{1/2}F=4\rightarrow 6P_{1/2}$ resonance by ~1/2 the ground-state hyperfine splitting $\nu_c$, which allows roughly equal scattering rates from both ground hyperfine levels. (We assume a modest bias magnetic field along the pump laser propagation direction). Raman scattering of the $\sigma^+$ pumping light drives $|F,m\rangle \rightarrow |F',m+p\rangle$ with $p=0,1,2$. For $p=1,2$ the mean value of $m$ is increased, so that repeated Raman scattering moves the state populations towards the fully electron and nuclear-spin polarized $|4,4\rangle$ state.


Of critical importance is the spectrum of the scattered light.  In the absence of collisions, the process proceeds entirely by scattering through the virtual state, shown dashed in Fig.~\ref{fig:CsRaman}a, so the scattered photon frequencies are $\nu_{sc}=\nu_0+(F-F')\nu_c+\Delta\boldsymbol{\kappa}\cdot\mathbf{v}$, where $\Delta\boldsymbol{\kappa}\cdot {\bf v}\sim  $0.3 GHz (400 K) is the Doppler shift of the scattered light. Thus the spectrum of the scattered light, shown in Fig.~\ref{fig:CsRaman}b is a convolution of a relatively narrow Doppler profile, appropriately averaged for the angular distribution of the scattering, with the primary Raman shifts $(F-F')\nu_c$.  It is critical that the pumping laser frequency be chosen so that all 3 Raman peaks have high escape probabilities.

\begin{figure*}[t]
\centering
\includegraphics[width=0.45\textwidth]{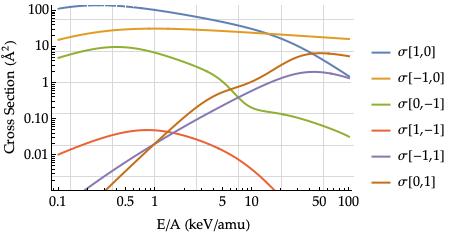}\hspace{0.04\textwidth}%
\includegraphics[width=0.45\textwidth]{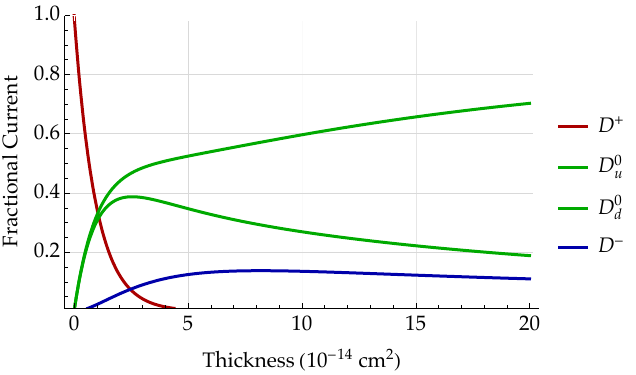}\hspace{0.04\textwidth}%
\caption{Charge-transfer pumping at high field. (a) Hydrogen-isotope charge-transfer cross sections on Cs from Ref.~\cite{Tatsuo1988}. (b) For 1 keV beams, polarized Cs suppresses the D$^-$ current because negative-ion formation is spin selective; this gives an experimental diagnostic of D$^0$ electron polarization. }
\label{fig:charge}
\end{figure*}

We turn now to the spin-relaxation mechanisms that set the scale for the required Raman scattering rates and consequently the required laser intensities.  Cs-Cs spin-exchange collisions will occur at a rate of about $10^5$/sec, but these conserve the total angular momentum and simply serve to promote a spin-temperature distribution of the Zeeman sublevels. The most fundamental angular momentum loss is the relaxation due to charge-transfer processes. The charge-transfer cross sections are shown in Fig.~\ref{fig:charge}.  The dominant neutralization process rapidly converts the incident positive ion beam into a neutral/negative mixture, so that the spin-relaxation is dominated by exchange between the neutral and negative ion currents, which is strongly polarization dependent. We may therefore assume that the Raman pumping has to compete with the formation of negative ions, with a cross-section $\sigma[0,-1]\sim 10^{-15}$ cm$^2$ (Fig.~\ref{fig:charge}). For a $\Phi=10$ Ampere beam with a $d=10$ cm diameter, we estimate an  relaxation rate for negative ion formation with an unpolarized D$^0$ current of
\begin{eqnarray}
    \Gamma_I=\sigma[0,-1]\Phi/A=800/{\rm sec}
\end{eqnarray}
The resulting Cs$^+$ ion, with its polarized Cs nucleus, will likely neutralize at the cell wall.  We presume that the remaining nuclear spin polarization is lost at the wall, so that each negative ion formation reaction fully relaxes the nucleus.

Neglecting radiation trapping, simulations \cite{HJW} of the Cs total angular momentum evolution in the spin-temperature limit can be described by
\begin{eqnarray}
\frac{d \langle F_z\rangle}{dt}&=&\frac{R}{3}\left(1-P\right)-\Gamma_I (1-P_D P) \langle F_z\rangle
\label{eq:dFdt}
\end{eqnarray}
Here the electron spin-polarization is $P=2\langle S_z\rangle$ and the photon scattering rate at zero polarization is $R$. The factor of 3 is exact in the absence of hyperfine structure, but modeling suggests it remains a good approximation when hyperfine interactions are included. We have also accounted for the reduction in the negative ion formation rate as the electron spin-polarization $P_D$ of the D$^0$ current increases. 

The steady-state polarization balance is
\begin{eqnarray}
P&=&\frac{R}{R+\frac{3}{2} q(P)\Gamma_I(1-P_DP)}
\label{eq:polbalance2}
\end{eqnarray}
where $q(P)=\langle F_z\rangle/\langle S_z\rangle$ accounts for the partitioning of angular momentum between electron and nuclear spin.  In the spin temperature limit \cite{Walker1997Spin-exchangeNuclei}, it ranges from 22 at small polarization to 8 as $P\rightarrow1$.
To reach a polarization of 0.8, using $q(0.8)=9.4$, requires $R=45,000$/sec for unpolarized $D$.  We emphasize that we have assumed the worst-case scenario, that charge transfer causes loss of the total Cs angular momentum, not just the electron spin.  We note that charge neutralization $D^-+Cs^+\rightarrow D+Cs^*$ should have a large cross section and could reduce the loss of Cs nuclear spin polarization.

To achieve a Cs polarization of 0.8 requires, assuming the relaxation is dominated by the fundamental ionization rate from the D beam,
\begin{eqnarray}
 {\cal P}=h\nu_0{R d^2\over\sigma(\Delta)}= 3.3\, {\rm kW}
\end{eqnarray}
using $\sigma=(\lambda^2/2\pi)A^2/4\Delta^2=3\times 10^{-16}$ cm$^{2}$, and the natural linewidth is $A=4.5$ MHz.  The dissipated power in a single pass through the vapor is much less:
\begin{eqnarray}
 {\cal P}_{\rm diss}=h\nu_0{R n d^3(1-P)}= 0.2\, {\rm kW}
\end{eqnarray}
The fact that the overall power requirement is much greater than the dissipated power per pass is consistent with the assumed small optical depths at the Raman detuning, and also implies that the power can be repurposed ~10 times to make a meter long cylinder for charge exchange polarization of the D neutral beam, as suggested in Fig.\ref{fig:Schematic}.  Alternatively, one could utilize a multi-pass cell to reduce the pumping inefficiency for a single 10 cm pumping volume.

Let us summarize the requirements for the laser.  Beyond the raw power, it is essential that its spectrum be within the transparency window of Fig.~\ref{fig:CsRaman}b, and in particular that it have negligible power near the optical resonances.  It is possible that one could get away with using a broadband Cs laser, then pass it through an optically thick unpolarized vapor cell to bleach the resonant portions of the spectrum.  Otherwise, the spatial mode quality of the laser is not particularly important as long as it reasonably fills the cell; rapid diffusion will spatially homogenize the Cs polarization.

In order to keep laser power requirements to a minimum, it is essential to control other sources of spin-relaxation. We begin with wall relaxation and proceed to radiation trapping.    
If the vacuum chamber walls are uncoated, we may assume they fully relax the Cs spins, electron and nuclear.  The mean-free path of the Cs atoms is $\lambda=1/n\sigma\sim 0.25$ cm, implying a diffusion coefficient of $D=v\lambda/3=2200$ cm$^2$/sec, and the consequent fundamental diffusion mode will produce an effective relaxation rate of at least
\begin{eqnarray}
    \Gamma_D\sim q D (2\pi^2)/d^2=4,000/{\rm sec} 
\end{eqnarray}
plus contributions from higher diffusion modes.

The wall relaxation problem may be effectively managed by using anti-relaxation coatings \cite{Seltzer2010InvestigationTechniques,Chi2020AdvancesCells}.  In particular, organochlorosilane (OTS) coatings are known to have an operating temperature of 170C that substantially exceeds the 135C needed for a 10$^{14}$/cc Cs density.  The results of Seltzer and Romalis \cite{Seltzer2009High-temperatureCoatings} for Rb suggest that it may be possible to achieve $b=100$ bounces before relaxation occurs, which would render the wall relaxation negligible compared to the fundamental relaxation from  charge transfer collisions.  

We return now to the issue of radiation trapping.  As seen from Fig.~\ref{fig:CsRaman}, the radiated light consists of 3 angular momentum components, and 3 frequency components.  Each angular momentum component has an angular distribution that resolves into a mixture of local polarization components when transported at a given angle through the atom cloud.  We therefore assume that spin polarization rises solely from the ordered pumping light.  Scattered photons that are re-absorbed before leaving the pumping region are effectively a relaxation mechanism.

Despite the  unpolarized nature of the scattered light, it has the very desirable feature that the dark state is preserved; as the atoms become highly polarized the number of scattered photons is reduced, reaching zero for fully polarized atoms.  Let us then consider this simplified model of the effects of radiation trapping:  the atoms undergoing Raman pumping scatter photons at a rate $R(1-P)$, where $P$ is the electron spin polarization.  These photons have an escape probability of $1-\epsilon$, so the re-absorbed photons can be considered to be an electron randomization process $d\langle F_z\rangle /dt=-\epsilon R(1-P)\langle S_z\rangle$, to be added to Eq.~\ref{eq:dFdt}, giving
\begin{eqnarray}
P&=&\frac{R}{R+\frac{3}{2}\epsilon R(1-P)+\frac{3}{2} q(P)\Gamma_I(1-P_DP)}
\label{eq:polbalance}
\end{eqnarray}
The power required to reach 80\% polarization for $\epsilon=0.2$ is thereby increased by about $4\epsilon (1-P)\approx 16\%$, to 
3.8 kW. The off-resonant Raman pumping allows the relaxation effects of radiation trapping to be manageable.

In the above estimation, we have ignored the stimulated Raman absorption probability for the scattered photons due to the presence of the pumping light. While the scattered photons are detuned from the excited level and therefore have high escape probability due to single-photon absorption (as shown in Fig.~2), the frequency difference between the scattered photons and the pumping light is two-photon resonant with the respective Raman transitions. As a result, resonantly enhanced two-photon Raman absorption cross section also needs to be taken into account. It is known that over a narrow frequency range of scattered-photon frequencies, the stimulated Raman absorption cross section can be as large as the resonant cross section of the hot vapor \cite{Yavuz2007ElectromagneticallyPulses}. This means that, a narrow set of scattered photons can experience the full optical thickness of the cloud, which can be as large as $n (\lambda^2 /2 \pi) (A/\Delta_{Doppler}) L \sim 5 \times 10^3 $. However, as we discussed above, the scattered photons are broadband and have a frequency spread of the Doppler width. As a result, the frequency-integrated absorption cross section over the two-photon Raman transition is the relevant quantity, which is reduced by a factor of $\sim (\Omega /2 \Delta ) ^2 $ compared to the resonant cross section of the hot vapor. Here, the quantity $\Omega$ is the Rabi frequency of the pumping light.  

For the required optical power of the pumping light (3.8~kW), the effective optical depth due to two-photon Raman absorption experienced by the scattered light is not negligible and is about $ (\Omega /2\Delta ) ^2 \times n (\lambda^2 /2 \pi) (A/\Delta_{Doppler}) L ~\sim 5 $. We note that while this suggests that stimulated Raman absorption is significant, it likely is not as detrimental as single-photon absorption of traditional radiation trapping. This is because, each stimulated Raman absorption event of the scattered light results in emission of a photon into the pumping laser (i.e. amplification), which can then continue to contribute to the optical pumping process. Furthermore, the back-of-the envelope estimate that we discuss here does not take into account (1) the inherent multi-level structure of the problem, and (2) the reduction in the Raman absorption cross section due to random direction and polarization of the scattered photons. A detailed analysis of radiation trapping under the conditions of Raman absorption in high-density multi-level alkali vapors will be among our future investigations.

\textit{Charge-transfer pumping--}
The charge dynamics of repeated charge-transfer collisions is determined by six processes, the most important being (in order of their magnitude at 0.5 keV/amu):
\begin{eqnarray}
    D^++Cs&\xrightarrow{\sigma[1,0]}& D^0+Cs^+ \label{eq:chargeexchange}\\
    D^-+Cs&\xrightarrow{\sigma[-1,0]}& D^0+Cs+e\\
    D^0+Cs&\xrightarrow{\sigma[0,-1]}& D^-+Cs^+ \label{eq:negform}
\end{eqnarray}
The cross-sections $\sigma[q_i,q_f]$ for these, plus the much weaker $\sigma[-1,0],\sigma[1,-1],\sigma[-1,1]$ processes, are summarized by \cite{Tatsuo1988} and shown in Fig.~\ref{fig:charge}.
We model these reactions using the methods of open quantum systems, giving a systematic way to generate the rate equations according to each process, while accounting for the various spin and nuclear spin effects. 

The large neutralization cross-section $\sigma[1,0]$ (\ref{eq:chargeexchange}), compared to the much smaller reverse processes $[0,1],[-1,1]$, results in essentially full conversion of an incident positive ion beam into neutrals and negative ions after just a few cm in the dense Cs vapor.  An important nuance of the [0,1] process is that the final state is mostly $n=2$, having close to the same binding energy as neutral Cs.  Any 2p atoms quickly radiatively decay, while at high current densities the 2s atoms likely also radiatively decay by Stark mixing with 2p. The reverse process [0,1] has a much larger energy defect because of the much larger binding energy of the H/D 1s state relative to Cs.

The negative ion fraction in unpolarized Cs equilibrates somewhat more slowly to approximately $\sigma[0,-1]/(\sigma[0,-1]+\sigma[-1,0])=0.24$ \cite{Schlachter1969Charge-ExchangeKevs,Schlachter1980DTargets}, as shown by the dashed curves in Fig. \ref{fig:charge}.

The collision dynamics are modified for charge transfer reactions in spin-polarized vapors. The [1,0] charge transfer is in principle spin-conserving, so that in fully polarized vapor the electron spin of the Cs is fully transferred to the largely n=2 neutral atoms.  This effect is well-known in optically pumped polarized ion sources \cite{Clegg2001ATechnology}. More importantly, since the only stable negative ion state a spin singlet, the [0,-1] negative ion formation of Eq.~\ref{eq:negform} generalizes to the two processes
\begin{eqnarray}
    D_u^0+Cs\uparrow&\xrightarrow{(1-P)\sigma[0,-1]}& D^-+Cs^+ \label{eq:negformup}\\
    D_d^0+Cs\uparrow&\xrightarrow{(1+P)\sigma[0,-1]}& D^-+Cs^+ \label{eq:negformdown}
\end{eqnarray}
for the spin-up and spin-down neutral atom currents.  Note that the cross-sections for the two channels are multiplied by polarization dependent factors $1\pm P$ that account for the probability that the collision proceeds through the singlet channel.

In addition to accounting for the spin-dependence of the collision processes, understanding the role of the magnetic field is also key.  There are two magnetic field scales in the problem.  The smallest is Breit-Rabi magnetic decoupling of the $D^0$ hyperfine structure, which occurs at fields above a few hundred Gauss.  For such fields, but smaller than the ~1 T fields needed to decouple the D$^0$ 2p fine structure, the radiative decay following the [0,1] charge transfer process into the n=2 levels causes partial loss of spin-polarization.  Even so, the $D^0$ currents become spin-polarized due to the charge transfer pumping effect.  The solid curves in Fig.~\ref{fig:charge}, which assume complete depolarization by the radiative decay,  show that not only does the neutral atom current become spin-polarized, but the negative ion current is suppressed.  The suppression is essentially complete in fully polarized Cs vapor, showing that the negative ion current is a sensitive measure of the neutral atom spin polarization and can be used as a useful diagonostic.

The second magnetic field scale is on the order of 1T, sufficient to decouple the 10 GHz 2p fine-structure splitting so that the 2p-1s spontaneous emission becomes spin-conserving.  In this case the polarizations attainable reach the Cs limit faster than shown in Fig.~\ref{fig:charge}. These calculations suggest a possible generalization of the OPPIS to high currents.  The output of a 20 cm charge-exchange tube could be sent through a Sona device \cite{Sona1967AD-}, with a rapid magnetic field reversal to transfer the high electron spin-polarization to nuclear spin-polarization.  For $D^0$, this results in maximum $\langle I_z\rangle=2/3$ and tensor polarization $\langle I_{zz}\rangle=0$, absent additional RF transitions \cite{Clegg2001ATechnology,Engels2021DirectUnit}.

\begin{figure}
    \centering
    \includegraphics[width=1\linewidth]{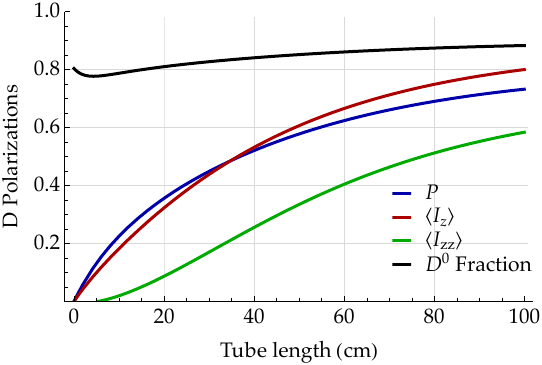}
    \caption{Generation of nuclear vector and tensor polarization by repeated charge transfer collisions in 90\% polarized Cs vapor.  }
    \label{fig:Dpols}
\end{figure}

We now turn to the charge transfer spin dynamics at low field, where hyperfine interactions allow the nuclear spins to become vector and tensor polarized along with the electrons, without Sona or RF manipulations.  This corresponds to the situation depicted in Fig.~\ref{fig:Schematic}.  The incident ion beam is first neutralized in 5 cm of unpolarized vapor, then the mixed D$^0$/D$^-$ beam proceeds through a long channel of polarized Cs vapor.  There are now 6 neutral atom currents, corresponding to the 6 hyperfine levels of the atoms, and 3 D$^-$ currents corresponding to the 3 nuclear spin states.  Our calculations include the 3 $D^+$ currents as well, though they play little role.  It is straightforward to generate the generalizations of Eqs.~\ref{eq:negformup} and \ref{eq:negformdown} using the magnetic field mixing of the various states, for each of the charge transfer processes.  Repeated charge transfer and hyperfine interactions polarize the D nuclei over a 1 meter path length, as shown in Fig.~\ref{fig:Dpols}.  

At the end of the charge transfer tube, the beam is largely nuclear and electron spin polarized neutral atoms.  At this point it may prove favorable to rotate the magnetic field to be longitudinal and $>1$ kG (not shown in Fig.\ref{fig:Schematic}) before a final charge transfer with unpolarized Cs to transform it into a 25\% $D^-$ beam, which can then be accelerated to the desired 200 keV scale energy.  Subsequent conversion back into a neutral atom beam at 200 keV will be necessary, perhaps by photodetachment \cite{Fiorucci2022OverviewReactors}.

\textit{Outlook--}
The above estimates suggest that it should be possible, with 5 kW scale Raman pumping of Cs, to generate a sufficiently long column of spin polarized Cs vapor to enable charge transfer pumping of nuclear spin polarized D at multiampere levels.  It should be emphasized that the potential 3-4 orders of magnitude increase over existing polarized ion source currents is only possible because the resulting atomic beam does not have to be delivered to a compact accelerator beam but instead to a ~0.5 m scale fusion plasma.

The ultimate efficiency of the process, conversion of positive $D^+$ into nuclear spin polarized $D^-$, should approach the 25\% charge transfer balance for 1 keV D atoms, assuming that the spreading of the neutral/negative ion beam in the charge transfer tube is manageable.  Assuming the Cs Raman light is generated by efficient nonlinear conversion from fibers lasers, or by highly efficient OPAL sources, the power requirements may be estimated, assuming 10\% efficiency, at 50 kW, 10\% of the 0.5 MW power of a 2.5 ampere, 200 keV negative ion beam.  

Beyond the myriad of practical details which need careful consideration into order to realize such a source, preliminary experiments demonstrating Raman pumping, performance of anti-relaxation coatings in the presence of ion beams, propagation of 1 keV $H^0/H^-$ beams through collisionally thick Cs vapors, and the effects of radiation trapping of Raman light are of the highest priority.  Recognizing these challenges, it nevertheless seems possible to generate multi-Ampere nuclear polarized atom beams for heating of fusion plasmas.


We greatly benefited from advice and suggestions from L. W. Anderson.

\bibliography{references}

\end{document}